\documentclass{article}

\usepackage{PRIMEarxiv}

\usepackage[utf8]{inputenc} 
\usepackage[T1]{fontenc}    
\usepackage{hyperref}       
\usepackage{url}            
\usepackage{booktabs}       
\usepackage{amsfonts}       
\usepackage{nicefrac}       
\usepackage{microtype}      
\usepackage{lipsum}
\usepackage{fancyhdr}       
\usepackage{graphicx}       
\graphicspath{{media/}}     

\usepackage{amsmath}
\usepackage{caption}

\usepackage{booktabs}
\usepackage{array}
\usepackage[numbers]{natbib}  

\title{Sentiment simulation using generative ai agents
}

\author{
\textbf{Melrose Tia}$^{1}$, 
\textbf{Jezreel Sophia Lanuzo}$^{1}$, 
\textbf{Lei Rigi Baltazar}$^{1}$, \\
\textbf{Marie Joy Lopez-Relente}$^{2}$, 
\textbf{Diwa Malaya Quiñones}$^{3}$, 
\textbf{Jason Albia}$^{1}$\thanks{Corresponding author: jason@netopia.ai} \\
$^{1}$Netopia AI, Inc., Manila, Philippines \\
$^{2}$Institute of Statistics, University of the Philippines Los Baños, Laguna \\
$^{3}$Department of Psychology, University of the Philippines Diliman, Quezon City \\
\texttt{\{melrose, sophia, lei, jason\}@netopia.ai, \{daquinones, mflopez2\}@up.edu.ph}
}

\begin{document}
\maketitle

\begin{abstract}
Traditional sentiment analysis relies on surface-level linguistic patterns and retrospective data, limiting its ability to capture the psychological and contextual drivers of human sentiment. These limitations constrain its effectiveness in applications that require predictive insight, such as policy testing, narrative framing, and behavioral forecasting. We present a robust framework for sentiment simulation using generative AI agents embedded with psychologically rich profiles. Agents are instantiated from a nationally representative survey of $2,485$ Filipino respondents, combining sociodemographic information with validated constructs of personality traits, values, beliefs, and socio-political attitudes. The framework includes three stages: (1) agent embodiment via categorical or contextualized encodings, (2) exposure to real-world political and economic scenarios, and (3) generation of sentiment ratings accompanied by explanatory rationales. Using Quadratic Weighted Accuracy (QWA), we evaluated alignment between agent-generated and human responses. Contextualized encoding achieved $92$\% alignment in replicating original survey responses. In sentiment simulation tasks, agents reached $81$\%–$86$\% accuracy against ground truth sentiment, with contextualized profile encodings significantly outperforming categorical ($p < 0.0001$, Cohen’s $d = 0.70$). Simulation results remained consistent across repeated trials ($\pm 0.2-0.5\%$ SD) and resilient to variation in scenario framing ($p = 0.9676$, Cohen’s $d = 0.02$). Our findings establish a scalable framework for sentiment modeling through psychographically grounded AI agents. This work signals a paradigm shift in sentiment analysis from retrospective classification to prospective and dynamic simulation grounded in psychology of sentiment formation. 

\end{abstract}


\section{Introduction}
Sentiment analysis involves assessing the opinions and attitudes toward specific areas of interests, playing a pivotal role in influencing decisions across business, societal, and individual domains \protect\protect\cite{pang2008opinion, liu2012sentiment}. While the term sentiment analysis gained prominence in the early 2000s \protect\cite{nasukawa2003sentiment, dave2003mining}, the broader practice of gauging public opinion has long shaped policy-making, democratic discourse, and marketing strategies \protect\cite{price1997opinion}. As digital platforms and user-generated content increasingly serve as channels for public expression, sentiment analysis enables organizations to harness opinion-rich and unstructured data to refine communication strategies and to respond effectively to societal trends.

In the socio-political domain, sentiment analysis has supported applications ranging from policy evaluation to campaign strategy by enabling large-scale interpretation of public opinion. Examples include assessments of public engagement with government initiatives \protect\cite{charalabidis2015opinion, sukma2020sentiment, hradec2023fables, vizmanos2023like}, political campaign analysis \protect\cite{sandoval2020sentiment, crabtree2020not, miranda2021exploring}, and citizen feedback monitoring via social media \protect\cite{umali2020sentiment}. For instance, Sandoval-Almazan et al. (2020) \protect\cite{sandoval2020sentiment} examined Facebook reactions to political campaign posts in Mexico, uncovering patterns in public engagement. In Indonesia, Sukma et al. (2020) \protect\cite{sukma2020sentiment} analyzed Twitter responses to the Omnibus Law, revealing levels of public support and dissent to the policy. In the Philippines, Miranda et al. (2021) \protect\cite{miranda2021exploring} tracked sentiment around presidential state addresses, while Umali et al. (2020) \protect\cite{umali2020sentiment} assessed citizen satisfaction with various government agencies based on social media commentary. 

Beyond politics, sentiment analysis is widely used in the private sector, where it serves as a critical tool in marketing, advertising, and customer experience strategies. Rathore et al. (2020) \protect\cite{rathore2020pre}, for example, analyzed emotional patterns in online comments before and after product launches to assess market reception and product fit. Giannakis et al. (2022) \protect\cite{giannakis2022social} showed how consumer sentiment from social media can inform early-stage product development, while Yin et al. (2022) \protect\cite{yin2022exploring} studied brand loyalty and satisfaction through Twitter sentiment toward e-commerce platforms Lazada and Shopee. In addition, sentiment analysis has also been applied to evaluate consumer reviews for predicting behavior and satisfaction \protect\cite{jain2021systematic, ghatora2024sentiment} and to generate real-time customer insights \protect\cite{krugmann2024sentiment}, thereby contributing to product refinement, enhanced customer engagement, and data-driven business strategies.

\subsection*{Traditional Sentiment Analysis and Their Limitations}

Traditional sentiment analysis often relies on structured methods such as surveys, opinion polls, and focus groups, alongside more recent digital sources like social media \protect\cite{liu2012sentiment}. These approaches have paved the way into computational techniques leveraging machine learning (ML) and natural language processing (NLP) to classify sentiment (e.g., negative, neutral, positive) based on large-scale text analysis. These methods analyze linguistic patterns, including the use of emotionally charged words (e.g., “happy”, “disappointed”) and syntactic structures that convey opinion or emotions.

Despite advances in ML and deep learning models that boost classification accuracy \protect\cite{mao2024sentiment}, these approaches are fundamentally limited. First, they primarily capture surface-level linguistic cues, often oversimplifying the complexity and nuance of human emotion and opinion. Second, these models function as black-box systems that lack transparency, offering limited insight into the reasoning behind sentiment predictions \protect\cite{jim2024recent}. This lack of interpretability impairs trust, accountability, and applicability in domains requiring nuanced understanding.

Third, and perhaps most critically, current sentiment analysis techniques often fail to account for contextual and psychological factors, including individual biases, personality traits, values, or temporal circumstances \protect\cite{mahmoudi2021identifying, lin2017personality, park2024personality}. For example, Mahmoudi (2021) \protect\cite{mahmoudi2021identifying} emphasizes how user-level biases can lead to divergent interpretations of the same event, which are often ignored in traditional models. Because these systems typically offer retrospective summaries rather than dynamic simulations, they struggle to support forward-looking applications such as policy testing, narrative impact studies, or synthetic focus groups \protect\cite{shrestha2025beyond}. 

To illustrate, a sentiment model trained on social media posts from a prior election may accurately classify political opinions from that period \protect\cite{chauhan2021emergence, khan2023improving}, but it cannot simulate how a specific group—such as rural, first-time voters, might react to a new policy announcement or media event. These limitations reveal a broader issue: these models are inadequate to model sentiment as situated cognition, that is, an emergent, psychologically grounded response shaped by internal dispositions and external stimuli \protect\cite{roth2013situated}.

\subsection*{Sentiment Simulation using AI and Behavioral Science}
Rooted in the above challenges, we propose a conceptual shift: from retrospective sentiment classification to AI and behavioral science-driven sentiment simulation. This approach integrates two core paradigms: (1) a behavioral science framework that explains how sentiments arise from psychological drivers, and (2) a simulation-based modeling paradigm enabled by generative AI.

Behavioral science provides the theoretical foundation for this shift. It conceptualizes sentiment as a dynamic construct shaped by cognition, emotion, and situational context. Social psychology suggests that sentiment reflects attitudes formed from beliefs, values, and environmental factors—factors that, in turn, shape behavior \protect\cite{myers2020ebook}. A complementary analysis by Li and Hovy (2017) \protect\cite{li2017reflections} further argue that sentiment originates from emotionally driven preferences and the pursuit of personal goals. These perspectives suggest that sentiment is not just a textual artifact but a behavioral expression rooted in individual psychology.

In methodical perspective, unlike traditional models that classify past sentiment, generative models such as large language models (LLMs) enable prospective simulations that can generate behaviorally rich, context-sensitive sentiment. These generative models can simulate trust dynamics \protect\cite{xie2024can}, personality expression \protect\cite{aher2023using}, and opinion formation \protect\cite{zhang2024llm}—capabilities that align well with psychological realism. In addition, generative models has also catalyzed new research on synthetic populations and simulated human studies \protect\cite{park2024genagents, guo2024llmsurvey}, positioning generative AI as a powerful tool for behavioral science. Representative studies illustrating these advances are summarized in Table \ref{table:studies}.

\begin{table}[htbp]
\centering
\caption{Recent studies that inform and support this work, highlighting their domains and key findings.}

\begin{tabular}{@{}>{\raggedright\arraybackslash}p{4.2cm} >{\raggedright\arraybackslash}p{3.2cm} p{7.5cm}@{}}

\toprule
\textbf{Study} & \textbf{Domain} & \textbf{Key Findings} \\
\midrule
Using LLMs to Simulate Multiple Humans and Replicate Human Subject Studies \protect\cite{aher2023using} & Behavioral
Economics and Social Psychology & Simulated classic behavioral studies (e.g., Ultimatum Game, Milgram) and found that larger LLMs (GPT-3.5/4) could replicate established findings across economics, psycholinguistics, and social psychology. \\
\addlinespace
Generative Agents: Interactive Simulations of Human Behavior \protect\cite{park2023simulcra} & Human-AI Interaction & Introduced "generative agents"—LLM-driven agents with memory, planning, and reflection. Demonstrated emergent behavior in interactive environments (e.g., autonomously organizing a Valentine’s Day party) from a single prompt. \\
\addlinespace
Generative Agent Simulations of 1000 People \protect\cite{park2024personality} & Social Science & Developed an LLM-based agent architecture to simulate 1,052 real individuals based on interviews. Agents replicated survey responses with $\approx 85\%$ accuracy, comparable to humans' own retest accuracy, and predicted personality traits well. \\
\addlinespace
User Behavior Simulation with LLM-based Agents \protect\cite{wang2023user} & User Behavior Simulation & Developed an LLM-based framework for simulating user behaviors (e.g., web navigation). Captured social dynamics like conformity and information cocooning. \\
\addlinespace
Can Large Language Model Agents Simulate Human Trust Behavior? \protect\cite{xie2024can} & Behavioral Economics & Used Trust Games to evaluate agent behavior. GPT-4 agents showed trust-like behavior and strong alignment with human responses in social dilemmas. \\
\addlinespace
Evaluating the Ability of LLMs to Emulate Personality \protect\cite{wang2025evaluating} & Personality Modeling & GPT-4 simulated individuals with Big Five profiles. Generated responses showed high internal consistency and strong correlation with self-reported personality scores. \\
\bottomrule
\end{tabular}
\label{table:studies}
\end{table}

While the above prior studies have illustrated potential of LLMs to simulate behaviors, replicate human experiments, or model trust, none have yet grounded sentiment simulation in real psychographic survey data. Our work fills this gap by embedding psychologically validated profiles into generative AI agents to simulate how real people might respond to socio-political and economic scenarios. 

\subsection*{Contribution of the Article}
In this study, we present a simple and scalable generative AI agentic framework via structured LLM prompting to simulate the sentiment response of the survey respondents on several socio-political and economic scenarios. The AI agents were instantiated to embody the psychological profiles derived from nationally representative survey and their simulated response is compared with the ground truth data. More precisely, the contributions of this work are as follows: 

\begin{itemize}
    \item We demonstrate that AI agents can be effectively instantiated to embody the psychological profiles constructed from empirically generated data. These profiles incorporate socio-demographic data and variables from validated psychological frameworks and attitudes on key socio-political and economic issues, providing agents with psychographically grounded priors.

    \item We show that these AI agents are capable of replicating survey results, as well as sentiment distributions observed in real-world responses, achieving high levels of individual-level alignment. Furthermore, we demonstrate that agent responses are robust across alternative framings of the same scenarios, indicating the consistency and stability of our simulation framework.
\end{itemize}

\setcounter{footnote}{0}

\section{Methodology}
\label{methods sec}

\subsection{Survey Design and Data Collection}
The survey instrument was designed to provide an interdisciplinary understanding of Filipino citizens' profiles by integrating multiple well-established psychological frameworks to capture a deeper understanding of public sentiment towards various socio-political and economic issues in the Philippines.

The instrument consists of $150$ items, integrating both sociodemographic variables (age, sex, educational attainment, religion, and other key identifiers) and different psychological dimensions (personality traits,  values, attitudinal frameworks, beliefs, and social and political behavior). These frameworks are theoretically grounded and considered temporally stable \protect\cite{apa_personality, russo2022changing, ye2024social}, allowing for the abstraction of consistent psychographic profiles. For greater sensitivity in capturing the intensity and direction of respondents’ responses, most frameworks were measured using a 7-point Likert scale. Respondents were asked to express their level of agreement or disagreement with statements about selected major socio-political and economic issues \protect\cite{pulseasiasona2024}. 

Descriptive statistics of a nationally representative sample of $2,485$ registered Filipino voters with $95$\% confidence level and $1.97$\% margin of error are summarized in Table \ref{table:sample_stat}. The respondents' age ranged from $18$ to $89$ years old, with the majority ($33$\%) falling within the adult age group ($28-42$ years old). The sample was gender-balanced ($50$\% female, $50$\% male), and the majority were married ($57$\%). In terms of socioeconomic status, nearly half of the sample ($49$\%) reported no monthly income, while $30$\% were categorized as low income. Most participants had completed at least high school ($52$\%) or college ($21$\%).

\begin{table}[htbp]
  \centering
  \caption{Descriptive statistics of the study sample $(N = 2,485)$.}
  \label{tab:descriptive_sample}
  \begin{tabular}{@{}llr@{}}
    \toprule
    \textbf{Variable} & \textbf{Category} & \textbf{Count (Relative Proportion)} \\
    \midrule
    \multicolumn{3}{@{}l}{\textbf{Age Group}} \\
    & Young Adults (18--27 Years Old)        & 399 (16\%) \\
    & Adults (28--43 Years Old)              & 820 (33\%) \\
    & Middle-Aged Adults (44--59 Years Old)  & 736 (30\%) \\
    & Seniors (60+ Years Old)                & 530 (21\%) \\
    \addlinespace
    \multicolumn{3}{@{}l}{\textbf{Marital Status}} \\
    & Single       & 380 (15\%) \\
    & Live-In      & 395 (16\%) \\
    & Married      & 1413 (57\%) \\
    & Separated    & 77 (3\%)   \\
    & Widowed      & 220 (9\%)  \\
    \addlinespace
    \multicolumn{3}{@{}l}{\textbf{Monthly Income-Based Socioeconomic Status}} \\
    & No Income     & 1213 (49\%) \\
    & Low Income    & 740 (30\%) \\
    & Middle Income & 530 (21\%) \\
    & High Income   & 2 ($<$1\%) \\
    \addlinespace
    \multicolumn{3}{@{}l}{\textbf{Highest Educational Attainment}} \\
    & No Formal Education     & 8 ($<$1\%)   \\
    & At least Elementary     & 502 (20\%)   \\
    & At least High School    & 1294 (52\%)  \\
    & At least Vocational     & 152 (6\%)    \\
    & At least College        & 525 (21\%)   \\
    & At least Graduate Studies & 4 ($<$1\%) \\
    \bottomrule
  \end{tabular}
  \label{table:sample_stat}
\end{table}

To our knowledge, our data represents the largest and most demographically diverse samples in the Philippines used to examine psychological frameworks, offering a robust basis for generalizing the findings to the broader adult population. Previous psychological studies in the Filipino samples, such as those by Church et al. (1997) \protect\cite{church1997filipino} $(N = 629)$, Del Pilar (2017) \protect\cite{delpilar2017development} $(N = 576)$, and Wapaño (2021) \protect\cite{wapanopersonality} $(N = 828)$, were conducted with smaller, more localized samples. 

\subsection{Sentiment Simulation}
\begin{figure}[htbp]
  \centering
  \includegraphics[width=0.62\textwidth]{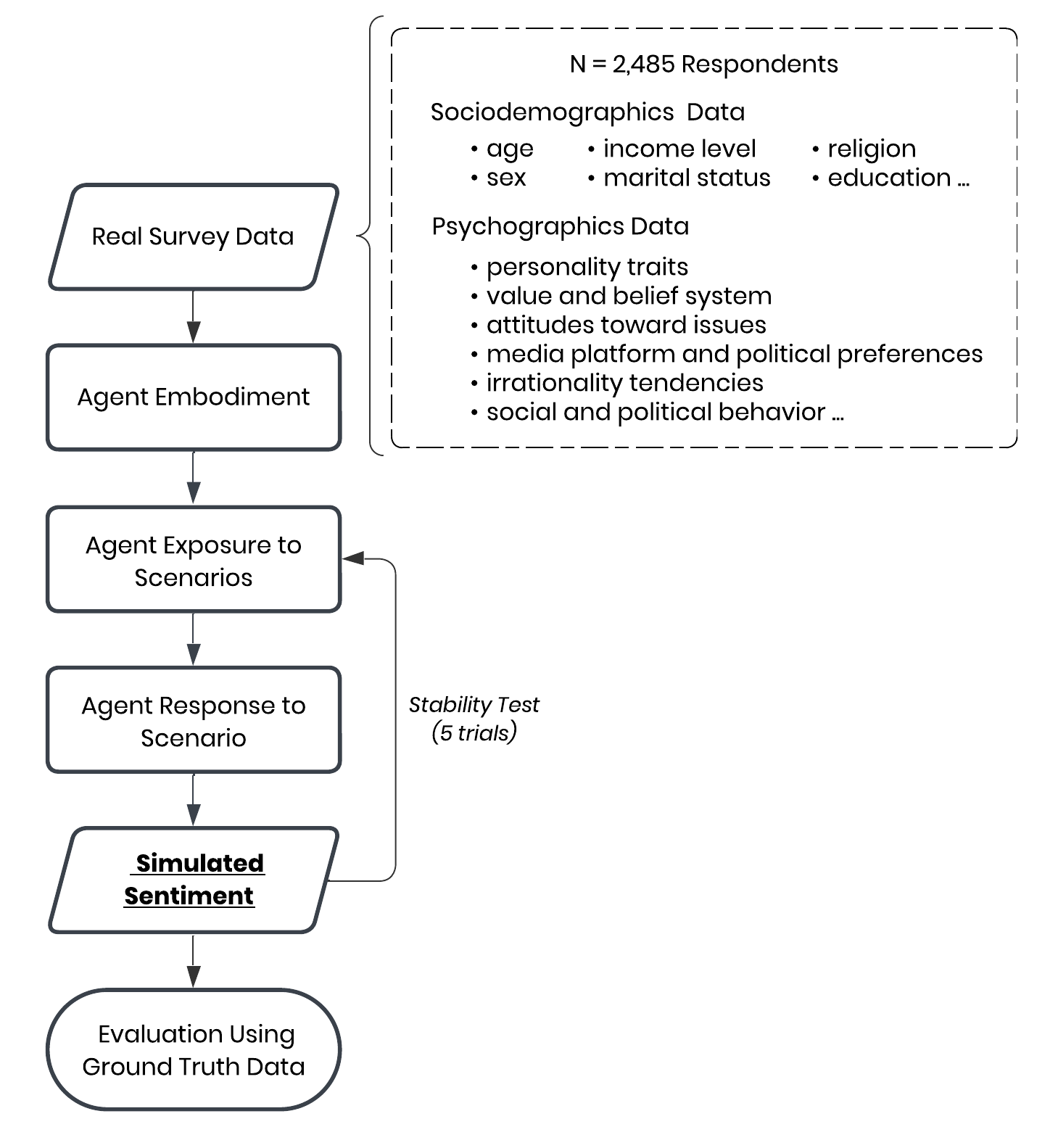}
  \caption{Sentiment Simulation Framework Using AI Agents.}
  \label{fig:sim_framework}
\end{figure}

The sentiment simulation framework leverages generative AI agents, embodied with psychographic and contextual variables, to model the sentiment of respondents in response to varying socio-political and economic scenarios. The framework enables generative agents to produce dynamic sentiment responses that are not only reactive to input stimuli but also aligned with their internal psychological attributes and contextual stimuli. As shown in Figure~\ref{fig:sim_framework}, the simulation framework consists of three (3) core stages: Agent Embodiment, Agent Exposure to Scenarios, and Agent Response to Scenarios.

All simulations were conducted using Llama 3.1 70B \footnote{Llama 3.1 70B was selected following rigorous experimentation with various LLMs evaluating their sensitivity to political and linguistic bias. \protect\cite{manus_in_progress}}, a state-of-the-art open-weight LLM optimized for instruction following, long-context reasoning, and alignment with human intent. This model is well-suited for simulating agent behavior within psychological frameworks due to its architecture that supports multi-turn coherence and robust language understanding \protect\cite{grattafiori2024llama}.

\subsubsection{Agent Embodiment}
Each AI agent is embodied with a unique set of sociodemographic and psychographic variables derived from empirical survey. These variables were embedded into prompt templates using one of two encoding strategies: categorical or contextualized.

\begin{itemize}
    \item Categorical encoding involved assigning discrete labels (e.g., Low, Moderate, High) to each psychological variable, producing a structured but abstract representation of personality and attitudes.

    \item Contextualized encoding, by contrast, translated these categories into narrative descriptions that reflect how psychological variables might manifest in scenario-relevant contexts. For example, high openness in policy domain might be expressed as receptive to new policy ideas or prone to considering multiple perspectives.    
\end{itemize}

To evaluate the effectiveness of embodiment, we conducted a survey replication task wherein each agent, embodied with a specific respondent's profile, answered the same Likert-scale survey items as the human participant. This task assessed whether the agent could faithfully reflect the individual’s psychological profile through simulated responses.

\subsubsection{Agent Exposure to Scenario}
In this phase, agents were presented with real-world scenarios analogous to campaign messages, policy debates, economic developments, or media coverage of socio-political and economic issues: budget transparency, political dynasties, inflation, the justice system, and wage policies. These scenarios are crafted as narrative prompts designed to elicit affective, cognitive, and psychographically grounded responses, engaging the agent's internal dispositions.

In addition, to examine the impact of scenario framing effects, each scenario was presented with either positive or negative polarity, simulating ideological differences in real-world discourse (e.g., progressive vs. conservative perspective). Respondents were randomly assigned to one framing type, while ensuring equal distribution of framing across the entire sample population.

\subsubsection{Agent Response to Scenario}
Following scenario exposure, each agent produced a structured sentiment response, rated on a 5-point Likert scale (Negative, Slightly Negative, Neutral, Slightly Positive, and Positive), along with a brief explanatory rationale for its simulated sentiment.

After generating its initial sentiment, the agent was prompted with a self-assessment task, asking whether its response was logically consistent with its psychographic profile and the characteristics of the scenario (see Supplementary Material \ref{sec:agent_response}). This iterative validation step reinforced coherence and internal consistency within the simulated responses.

\subsection{Performance Evaluation Metrics}
\subsubsection{Quadratic Weighted Accuracy (QWA)}

QWA was employed as the primary metric to evaluate alignment between agent-generated and human responses on an ordinal scale. It penalizes distant misclassifications more heavily than near-miss errors, making it particularly suitable for Likert-scale classification tasks, where response categories are inherently ordered.

The QWA score is computed using Eq.~\eqref{eq:qwa_formula}, with weights that increase quadratically based on the distance between simulated and actual responses. This scoring method allows for a more nuanced assessment of model performance, rewarding response predictions that are close to the expected value even when they are not exact matches.

\begin{equation}
w_{ij} = 1 - \left( \frac{d_{ij}}{d_{\text{max}}} \right)^2
\label{eq:qwa_formula}
\end{equation}

where:
\newline
$w_{ij}$ is the score assigned to the pair of categories $i$ (true response) and $j$ (simulated response);
\newline
$d_{ij}$ is the absolute distance between the true and simulated response categories; and
\newline
$d_{\text{max}}$ is the maximum possible distance given the range of all possible response categories.

Higher QWA scores indicate that the agents’ responses are statistically accurate and internally coherent, i.e., interpretable within the context of their embodied psychological profiles. Score matrices are visualized in Supplementary Materials \ref{section:supp_embodiment_qwa} and \ref{section:supp_sentiment_qwa}.

\subsubsection{Statistical Tests}

To evaluate the statistical significance of observed differences in agent–human alignment, we employed both parametric (paired t-test) and non-parametric (Wilcoxon signed-rank) analysis, depending on the distributional properties of the QWA scores. Specifically, paired t-test was used when the assumption of normality was satisfied, whereas Wilcoxon signed-rank test was applied when this assumption was violated, due to their robustness to non-normal distributions. A commonly used threshold of $p<0.05$ was used to determine statistical significance.

In addition to hypothesis testing, we computed Cohen’s $d$ to estimate effect sizes and assess the practical relevance of observed differences. Effect sizes were interpreted using standard benchmarks: $d \approx 0.2$ (small), $d \approx 0.5$ (medium), and $d \geq 0.8$ (large). This dual approach enabled a robust interpretation ensuring that the reported improvements in alignment were not only statistically significant but also practically meaningful.

\section{Results and Discussion}
\subsection{Agent Embodiment Evaluation}

Agent embodiment was implemented using two distinct encoding strategies: categorical encoding, which uses ranked labels (e.g., Low, Moderate, High), and contextualized encoding, which embeds psychological variables into narrative descriptions. These strategies offer differing levels of abstraction in representing individual profiles, allowing us to compare their effects on simulated sentiment alignment. 

These encoding strategies draw from recent works that attempt to embed psychological traits into LLM prompts. For example, Wang et al. (2025) \protect\cite{wang2025evaluating} used personality assessment data, albeit limited to numeric Big Five scores, to prompt GPT-4 in simulating individual behaviors. Their method mirrors our categorical encoding approach, which also draws from empirical data but translates scores into ranked labels such as Low, Moderate, or High. In contrast, Xie et al. (2024) \protect\cite{xie2024can} used structured prompts with demographic and background details, similar to our contextualized strategy, to elicit trust behaviors from LLMs. Our study advances these efforts by grounding both encoding strategies in real large-scale survey data, allowing systematic comparisons between encoding levels.

Agent alignment with human survey responses is measured using QWA, where identical ratings yield $100$\% accuracy score and one-point differences result in proportionally lower score of $97$\%, capturing the degree of ordinal misalignment. See Supplementary Material~\ref{section:supp_embodiment_qwa} for details.

\begin{figure}[htbp]
  \centering
  \includegraphics[width=0.55\textwidth]{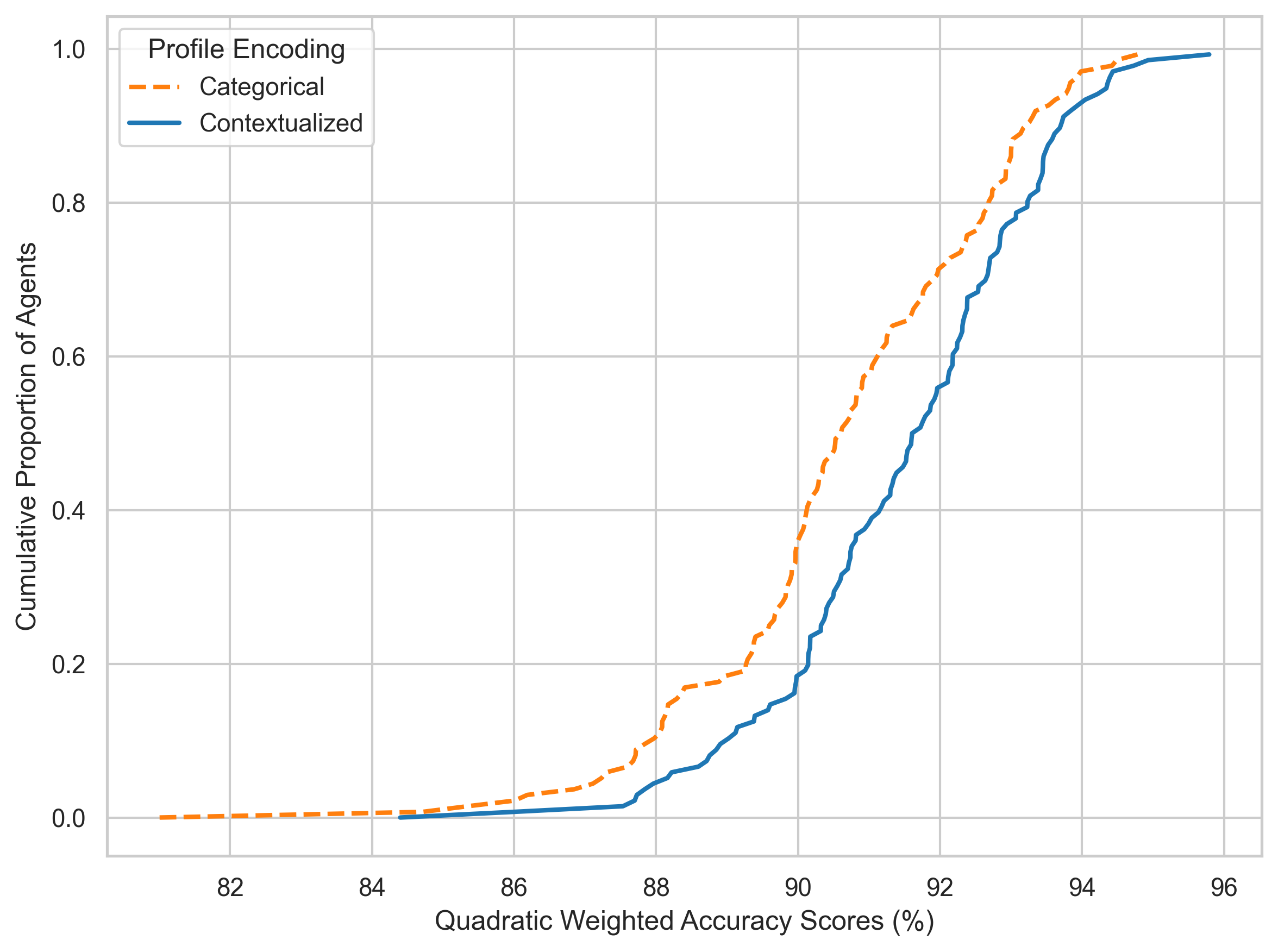}
  \captionsetup{width=0.6\textwidth} 
  \caption{Cumulative Distribution Function (CDF) Plot: Distributional Comparison of QWA Scores Across Profile Encoding Strategies.}
  \label{fig:cdf_plot}
\end{figure}

Figure \ref{fig:cdf_plot} illustrates the distribution of QWA scores for the two encoding strategies. The contextualized group’s curve (blue) is consistently right-shifted, indicating that a larger proportion of agents achieved higher alignment scores compared to their categorically encoded counterparts. This population-level trend suggests that narrative profile encoding enables more human-consistent responses.

\begin{figure}[htbp]
  \centering
  \includegraphics[width=0.55\textwidth]{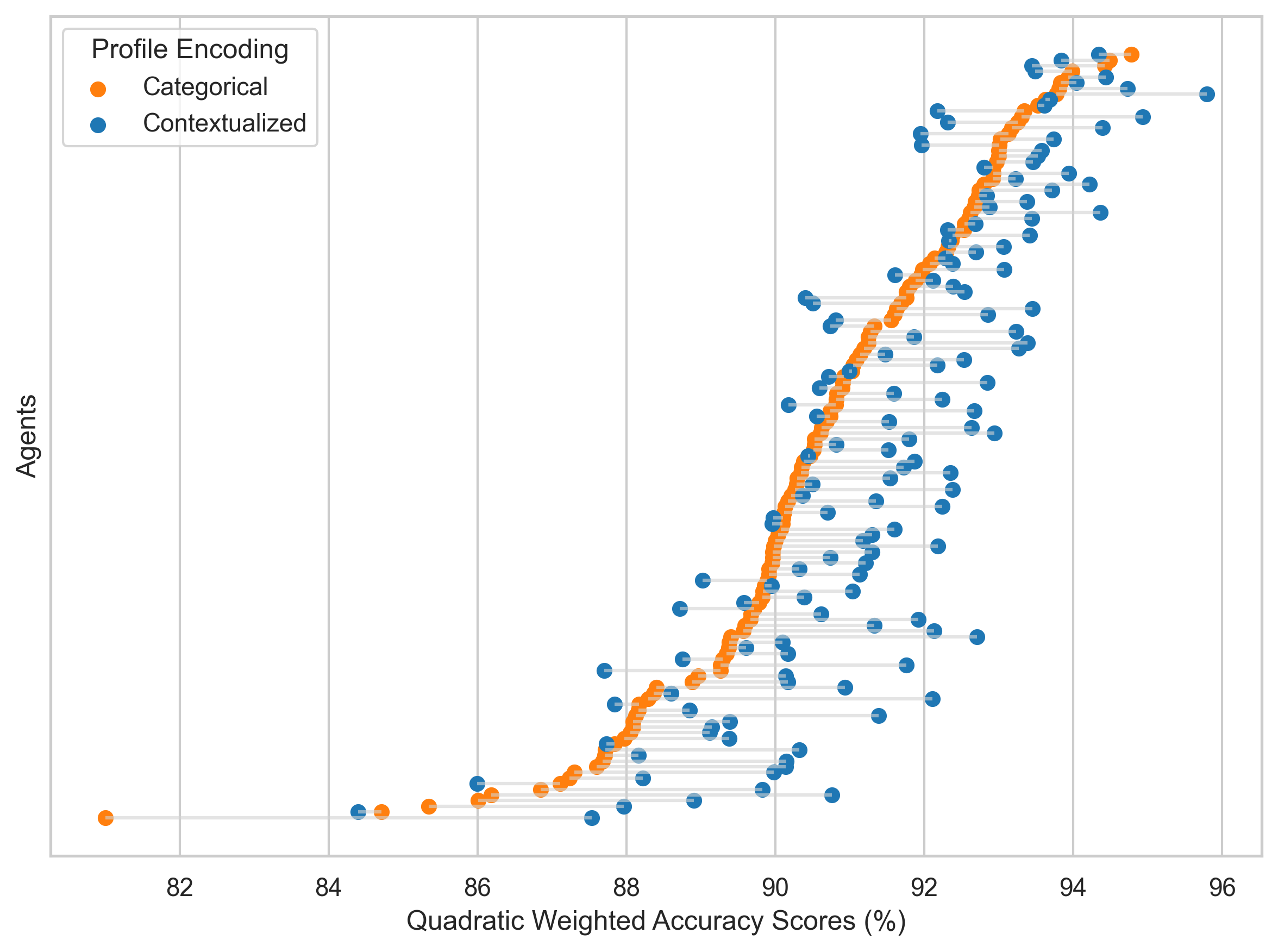}
  \captionsetup{width=0.7\textwidth} 
  \caption{Paired Dot Plot: Per-Agent Comparison of QWA Scores Across Profile Encoding Strategies. The vertical axis represents agents that are indexed arbitrarily.}
  \label{fig:dot_plot}
\end{figure}

Figure \ref{fig:dot_plot} offers an agent-level comparison. Each line connects the categorical and contextualized scores for a single agent, highlighting changes in alignment. Most lines extend rightward, reinforcing that contextualized encoding generally results in improved alignment for individual agents.

To determine whether the observed performance difference was statistically significant, we employed a Wilcoxon signed-rank test. Preliminary diagnostics using the Shapiro–Wilk test indicated violations of normality $(p = 0.0004$, justifying the use of a non-parametric approach. The Wilcoxon signed-rank test yielded a significant result $(p < 0.0001)$, suggesting that the alignment advantage of contextualized profile encoding is unlikely to be attributable to random variation. To assess the practical significance of this effect, we calculated Cohen’s $d = 0.70$, indicating a moderate effect size. Interpreted probabilistically, this reflects a $76$\% chance that a randomly selected agent with contextualized encoding would outperform one using categorical encoding in response alignment \protect\cite{mcgraw1992common}. These findings provide statistical and practical evidence that contextualized profile encoding yields better alignment with human responses compared to categorical encoding.

On average, agents using contextualized profiles achieved $92$\% alignment with original human responses, demonstrating the model's capacity to simulate individual-level psychographic data with high fidelity. These results compare favorably with prior efforts such as \protect\cite{kim2024llmmirror}, which introduced the LLM-Mirror framework to assess the consistency between LLM-generated responses and human survey data. While their persona-based prompting achieved $69$\% to $73\%$ consistency in domains like online advertising, corporate reputation, and customer loyalty, our approach reaches notably higher alignment levels across a broader array of psychological constructs. Similarly, Yeykelis et al. (2024) \protect\cite{yeykelis2024aipersona} found that AI personas could reproduce findings from experimental media studies with a $76$\% success rate. Our $92\%$ alignment suggests a stronger capacity to simulate nuanced attitudinal data, particularly when narrative context is used to express psychological variables. 

Collectively, these results demonstrate that contextualized psychological profile encoding significantly enhances agent-human alignment and produces more consistent responses. Contextualized encodings guide agents more effectively by embedding psychological traits within descriptive, scenario-relevant narratives. The performance gap between categorical and contextualized encodings highlights the benefits of translating psychological variable labels into rich psychographic contexts, enabling agents to respond more accurately in alignment with their profiles—a critical foundation for generating psychologically coherent sentiment simulations.

\subsection{Sentiment Simulation Performance}

Following the high alignment observed in the agent embodiment task, we next evaluate the ability of psychographically grounded agents to simulate human sentiment across a set of socio-political and economic scenarios: wage policies, budget transparency, inflation, the justice system, and political dynasties. This analysis provides a broader test of the model's ability to generate human sentiment responses in real-world contexts.

\begin{table}[htbp]
  \centering
  \caption{Sentiment Simulation Accuracy Across Socio-Political and Economic Scenarios.}
  \begin{tabular}{l@{\hspace{2em}}c c@{\hspace{2em}}c c}
    \toprule
    \textbf{Scenario} 
    & \multicolumn{2}{c}{\textbf{Categorical}} 
    & \multicolumn{2}{c}{\textbf{Contextualized}} \\
    & \multicolumn{1}{c}{\textbf{Average}} & \multicolumn{1}{c}{\textbf{SD}} 
    & \multicolumn{1}{c}{\textbf{Average}} & \multicolumn{1}{c}{\textbf{SD}} \\
    \midrule
    Wage Policies         & 80.3\% & $\pm$ 0.19\% & \textbf{83.4\%} & $\boldsymbol{\pm}$ \textbf{0.20\%} \\
    Budget Transparency   & 80.1\% & $\pm$ 0.21\% & \textbf{82.9\%} & $\boldsymbol{\pm}$ \textbf{0.33\%} \\
    Inflation             & 74.9\% & $\pm$ 0.32\% & \textbf{81.8\%} & $\boldsymbol{\pm}$ \textbf{0.17\%} \\
    Justice System        & 86.7\% & $\pm$ 0.39\% & \textbf{86.2\%} & $\boldsymbol{\pm}$ \textbf{0.26\%} \\
    Political Dynasties   & 68.4\% & $\pm$ 0.20\% & \textbf{81.2\%} & $\boldsymbol{\pm}$ \textbf{0.51\%} \\
    \bottomrule
  \end{tabular}
  \label{tab:sim_performance}
\end{table}

Table \ref{tab:sim_performance} summarizes sentiment alignment performance across the scenarios, comparing categorical and contextualized encoding strategies. As shown, contextualized encoding consistently outperformed categorical encoding in four out of five scenarios, with alignment accuracy gains ranging from $2.8$\% to $12.8$\% points. While categorical encoding achieved accuracy levels ranging from $68$\% to $87$\%, contextualized profile encoding yielded more stable and higher performance of $81$\% to $86$\%. 

The largest accuracy gain occurred in the political dynasties scenario ($+ 12.8\%$), followed by inflation ($+6.9\%$). For wage and budget transparency, improvements were more modest ($+2.8\%$ and $+ 3.1\%$, respectively). Interestingly, performance was nearly identical in the justice system scenario ($-0.5$\%), suggesting that some scenarios may be less influenced by internal psychological factors and more driven by ideological alignment or external cues.

These findings reinforce that sentiment simulation is enhanced when agents are grounded in contextually expressed psychological traits, not merely categorical summaries. The more realistically an agent’s internal disposition is modeled, the more accurately it mirrors human responses. This supports existing research \protect\cite{xie2024can} indicating that contextual richness improves behavioral realism in LLM simulations.

Considering the inherent variability of LLMs, stemming from prompt sensitivity and randomness introduced by stochastic decoding, we evaluated the stability of simulation outputs over repeated trials. Each scenario was simulated five ($5$) times, and performance was averaged to assess internal consistency. As shown also in Table \ref{tab:sim_performance}, sentiment alignment scores were highly stable, with standard deviations for contextualized encoding ranging from $\pm 0.17\%$ to $\pm 0.51\%$, indicating minimal variability in performance across trials. More precisely, the justice system scenario exhibited the highest and most stable performance, with QWA scores ranging narrowly from $86.0$\% to $86.7$\%. Wage policies and budget transparency also showed strong stability, with QWA scores clustered tightly around the mid-$83$\% range. Inflation followed a similar trend, with minor fluctuations around $82$\%. Although political dynasties had the lowest overall scores, ranging from $80.1$\% to $81.4$\%, the variation across trials was still minimal, indicating internal consistency even in comparatively more complex or ideologically loaded scenarios. 

Ultimately, our framework achieved high alignment performance across all tested scenarios ($81$\% to $86$\%), reflecting not only the predictive accuracy of the model, but also its behavioral plausibility. The framework’s consistency across trials is illustrative of its suitability for use in replicable and scalable behavioral simulations. Our findings highlight three pillars of effective simulation in behavioral science specifically in social sciences: (1) psychological grounding through contextualized traits, (2) consistency of performance across diverse and complex scenarios, and (3) sentiment alignment with empirically plausible human behavior \protect\cite{falk2009lab, lazer2009computational}. Moreover, in light of the variability inherent in emotional reasoning and the influence of framing on an individual's judgment \protect\cite{kahneman1984choices}, our results speak not only to technical performance, but to the psychological credibility of the simulated agents themselves.

\subsubsection{Simulation Robustness to Scenario Framing}
To further evaluate the framework’s generalizability, we investigated its sensitivity to framing effects, i.e., whether sentiment alignment varied substantially depending on whether a scenario was presented in a positive or negative light (e.g., performing well under positive framing but poorly under negative framing). This step is important given that prior studies in behavioral sciences and communication have shown that framing can substantially alter public opinion \protect\cite{chong2007framing, sniderman2004structure}.

\begin{figure}[htbp]
  \centering
  \includegraphics[width=0.58\textwidth]{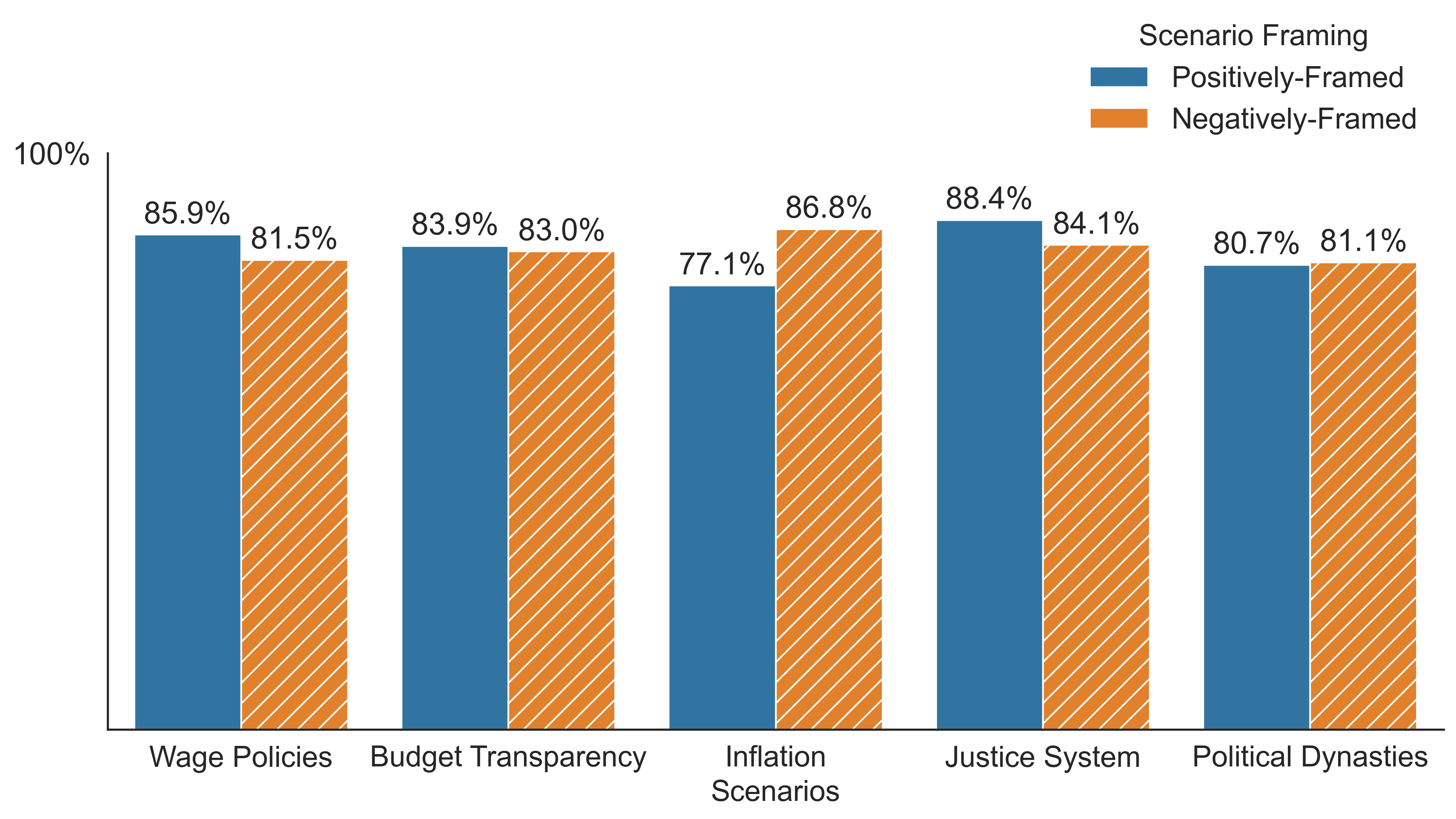}
  \captionsetup{width=0.7\textwidth}
  \caption{Quadratic Weighted Accuracy Between Survey and Simulated Sentiments Across Framing Types of the Different Scenarios.}
  \label{fig:framing}
\end{figure}

Figure \ref{fig:framing} shows a plot comparison between the average QWA for positive (blue) and negative (orange) framings for each scenario. Across the five socio-political and economic scenarios, QWA scores remained high $77$\% to $88$\%, with no consistent performance degradation or amplification due to framing. While differences between the positively- and negatively-framed scenarios ranged from $0.4$\% to $9.7$\%, the directionality and magnitude of these differences varied across scenarios. For example, negatively-framed scenarios yielded higher alignment in inflation ($+ 9.7\%$) and political dynasty topics $(+ 0.4\%)$, whereas positively-framed scenarios outperformed in justice system ($+ 4.3\%$), wage policies ($+ 4.4\%$), and budget transparency ($+ 0.9\%$). 

In addition, to further evaluate whether scenario framing influences sentiment simulation accuracy, we conducted a paired sample t-test comparing agent–human alignment scores across positively- and negatively-framed versions of each issue. The paired t-test was chosen to assess mean differences between framing conditions, with the Shapiro–Wilk test confirming that the normality assumption was sufficiently met  ($p = 0.1388$) . The analysis yielded a non-significant result ($p = 0.9676$), indicating no statistically meaningful difference in simulation accuracy across framing conditions. Furthermore, to quantify the magnitude of any potential effect, we computed Cohen’s $d = 0.02$, reflecting a negligible effect size. This suggests that the difference in QWA scores between framing conditions is practically insignificant, with sentiment alignment performance remaining stable regardless of scenario prompt framing.

Collectively, these results indicate that scenario framing does not exert a consistent or meaningful influence on simulation accuracy. The framework allows agents to anchor their evaluations to their  psychological  attributes, rather than being influenced by the differences in the scenario polarity framing.

These findings suggest that the agents remained anchored to their psychographic grounding, even under affective variation in scenario prompts. From a behavioral science perspective, this mirrors the consistency of human behavior across varied contexts, as documented in research on trait-based models \protect\cite{mccrae1997personality}. This coherence supports the notion that rich, context-sensitive embeddings enable psychologically grounded rather than context-reactive responses.

\section{Conclusion}
This study presents a psychographically grounded framework for sentiment simulation, leveraging language model agents embodied with empirically derived psychological profiles. By integrating validated constructs into structured prompts, we enable AI agents to simulate sentiment responses that are context-sensitive, psychologically coherent, and behaviorally plausible.

Our evaluation demonstrates that agents instantiated with contextualized profile encodings closely replicate individual-level sentiment patterns. In a survey replication task, these agents achieved alignment scores of up to $92$\%, significantly outperforming categorical encoding strategies. This result underscores the importance of narrative-rich representations in capturing the depth and nuance of human sentiment.

Beyond static replication, the framework also performs reliably in dynamic simulation tasks. When exposed to real-world socio-political and economic scenarios, agents achieved high alignment accuracies indicating their capacity to model realistic sentiment responses. Importantly, these results remained highly stable across five independent trials and different scenario framings, highlighting the internal consistency of the framework despite the stochastic nature of language models.

Overall, these results establish a reliable, scalable, and psychologically informed method for modeling public sentiment. The framework offers practical applications in policy testing, narrative framing analysis, and the development of synthetic populations for large-scale social simulation. More broadly, this work marks a paradigm shift—from retrospective sentiment classification toward prospective, psychologically grounded simulation leveraging the intersection of generative AI and behavioral sciences.

\section*{Acknowledgments}
We extend our sincere thanks to Mojhune Gabriel Manzanillo for his dedicated work in generating the experimental results for this study. We also gratefully acknowledge Adrian Gabonada for his insightful contributions, which significantly enriched the behavioral science interpretation and the discussion of our findings. We further thank Dannah Zemirah Junio for her guidance on statistical analysis; her input was instrumental in ensuring the rigor and validity of our evaluation methods. Model inferences and sentiment simulation were performed using compute resources provided by the Google Cloud for Startups Program.


\newpage

\newcounter{supplsec}
\newcounter{supplsubsec}[supplsec]
\renewcommand{\thesupplsec}{\Alph{supplsec}}
\renewcommand{\thesupplsubsec}{\thesupplsec.\arabic{supplsubsec}}

\newcommand{\supplsection}[1]{%
  \refstepcounter{supplsec}%
  \bigskip\noindent\textbf{\thesupplsec. #1}\par\nopagebreak
}

\newcommand{\supplsubsection}[1]{%
  \refstepcounter{supplsubsec}%
  \medskip\noindent\textbf{\thesupplsubsec\quad #1}\par\nopagebreak
}

\section*{Supplementary Material}
\addcontentsline{toc}{section}{Supplementary Material}

\supplsection{Survey Design and Implementation Details}

\supplsubsection{Survey Design and Instrument Specifics}
Specific psychological frameworks include  personality traits (e.g., HEXACO personality), values (e.g., Basic Personal Values), attitudinal frameworks (e.g., Affective Intelligence Theory), and beliefs (e.g., Social Axioms). It also encompasses social and political behavior (e.g., Civic Engagement). In addition to the sociodemographics and psychological frameworks, the survey instrument also includes an additional section to assess general citizen attitudes toward four major economic issues (e.g., inflation, minimum wage, etc) and four key social issues (e.g., the West Philippine Sea dispute, corruption, etc).

\supplsubsection{Survey Sampling}
A multi-stage stratified random sampling design was used to obtain a nationally representative sample of $2,485$ Filipino adults \protect\cite{comelec2022rvva}. The sample was proportionally distributed across the $17$ administrative regions using probability proportional to size. Systematic interval sampling selected five $(5)$ households per sampled barangay, and one $(1)$ respondent per household was randomly chosen using gender-rotated probability to ensure balanced male and female representation. This sampling design accounted for clustering at multiple geographic levels and stratification by region and urbanicity. Data were collected through face-to-face interviews from November $22$ to December $9$, $2024$.  A hybrid system of digital tablets and printed forms was used in the field to ensure both flexibility and high data fidelity.

\supplsubsection{Weighting Procedure}
To correct for unequal selection probabilities inherent in the sampling design, design weights (base weights) were computed from the joint probabilities of selection at each sampling stage: cities/municipalities, barangays, households, and eligible respondents. These base weights were then adjusted using post-stratification techniques, anchored on the official registered voter count data \protect\cite{comelec2022rvva} by region and gender. This procedure ensured that the final weighted sample reflected the actual distribution of registered voters, thereby improving the generalizability and precision of population-level inferences and estimates. 

\supplsubsection{Survey Implementation}
Randomization techniques were applied to minimize selection bias, and interval sampling was employed to ensure systematic coverage of both urban and rural areas. The use of in-person interviews allowed for greater engagement and clarification of questions when necessary, contributing to higher response quality and completeness.

\begin{figure}[htbp]
  \centering
  \includegraphics[width=0.56\textwidth]{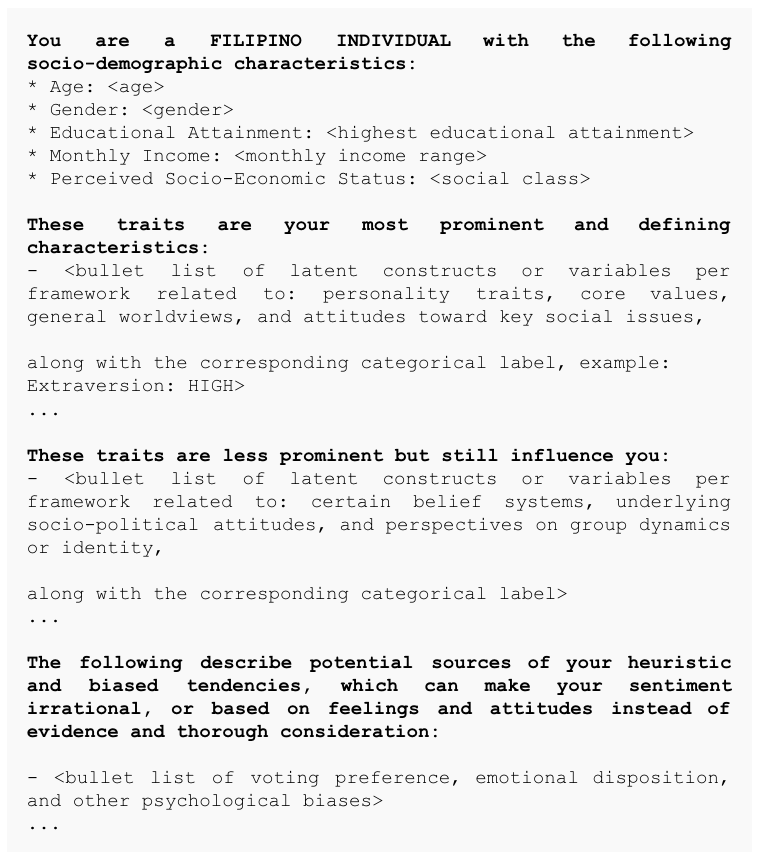}
  \captionsetup{width=0.7\textwidth}
  \caption{Prompt Format for Categorical Profile Encoding.}
  \label{fig:categorical}
\end{figure}

\begin{figure}[htbp]
  \centering
  \includegraphics[width=0.56\textwidth]{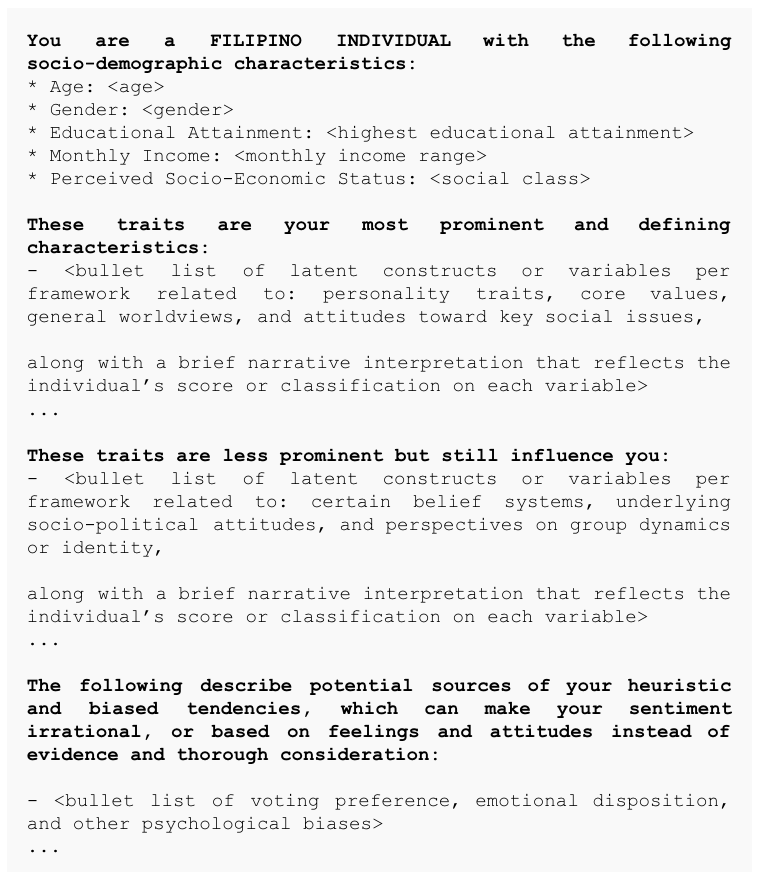}
  \captionsetup{width=0.7\textwidth}
  \caption{Prompt Format for Contextualized Profile Encoding.}
  \label{fig:contextualized}
\end{figure}

\supplsection{Agent Embodiment Setup}
Figures \ref{fig:categorical} and \ref{fig:contextualized} present the prompt formats used in the agent embodiment section of the sentiment simulation. Both formats operationalize the presentation of sociodemographic and psychographic attributes to the language model, serving as the foundation for generating agent-specific responses. The categorical format (Figure \ref{fig:categorical}) conveys traits and attributes through compact, labeled variables (e.g., Extraversion: HIGH), while the contextualized format (Figure \ref{fig:contextualized}) embeds the same information within brief narrative descriptions, enriching each variable with interpretive context.

In both cases, bracketed fields (e.g., $<age>$, $<income range>$) represent placeholders dynamically populated with real survey data during prompt instantiation. Text segments rendered in bold correspond to fixed prompt components that remain consistent across all agents.

\begin{figure}[htbp]
  \centering
  \includegraphics[width=0.58\textwidth]{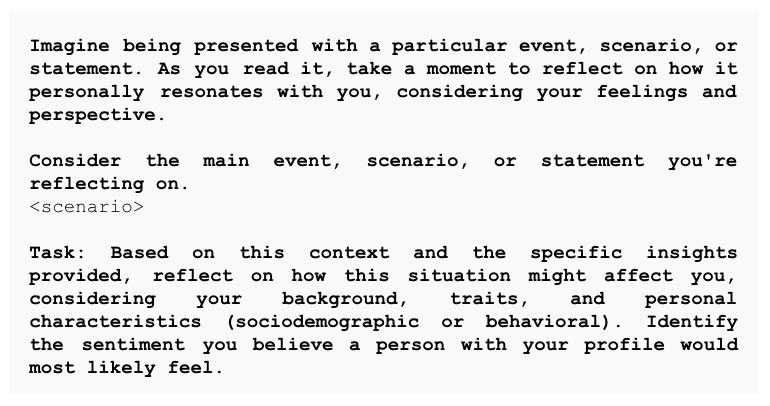}
  \captionsetup{width=0.7\textwidth}
  \caption{Prompt Format for Instantiating Agent Exposure to Scenario.}
  \label{fig:agent_exposure}
\end{figure}

\supplsection{Agent Exposure to Scenario}
Figure \ref{fig:agent_exposure} presents the prompt structure used to expose an embodied agent to a situational stimulus and elicit a corresponding affective judgment or sentiment response. In this format, the language model is prompted to imagine being presented with a particular event, scenario, or statement, and to reflect on how it would personally resonate based on the agent’s encoded background and perspective.

The $<scenario>$ placeholder is dynamically filled with the target stimulus, while all bolded text constitutes fixed instructional language consistent across all prompts. The model is then asked to identify the sentiment that best reflects how someone with its assigned profile would most likely feel in response.

\begin{figure}[htbp]
  \centering
  \includegraphics[width=0.58\textwidth]{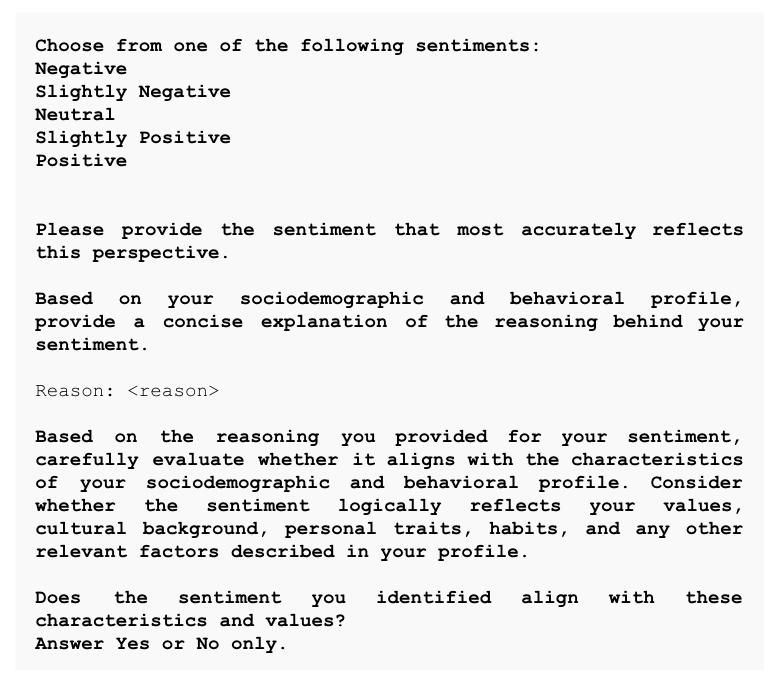}
  \captionsetup{width=0.7\textwidth}
  \caption{Prompt Format for Generating Agent's Response to Scenario.}
  \label{fig:agent_response}
\end{figure}

\supplsection{Agent Response to Scenario}
\label{sec:agent_response}
Figure \ref{fig:agent_response} presents the full instruction sequence used to elicit a sentiment judgment, accompanying rationale, and self-assessed alignment from an embodied agent profile. After being exposed to a scenario, the agent is instructed to identify the sentiment that most accurately reflects how a person with that profile would likely respond.

In addition to selecting a sentiment from a standardized 5-point scale (Negative to Positive), the model is prompted to articulate a brief explanation for its judgment. The $<reason>$ placeholder denotes the position where the model is expected to generate this response. Following this, the model is asked to critically evaluate whether its chosen sentiment logically aligns with the profile’s described characteristics, including values, personal traits, and contextual background, and to answer with a binary Yes or No. This prompt format supports deeper analysis of the model’s internal coherence, linking sentiment expression to reasoning and value alignment within an embodied simulation context. Similarly, all bolded segments represent fixed instructional text presented uniformly across prompts.

\supplsection{Quadratic Weighted Accuracy (QWA) as Evaluation Metric}
\label{sec:qwa}

\begin{figure}[htbp]
  \centering
  \includegraphics[width=0.63\textwidth]{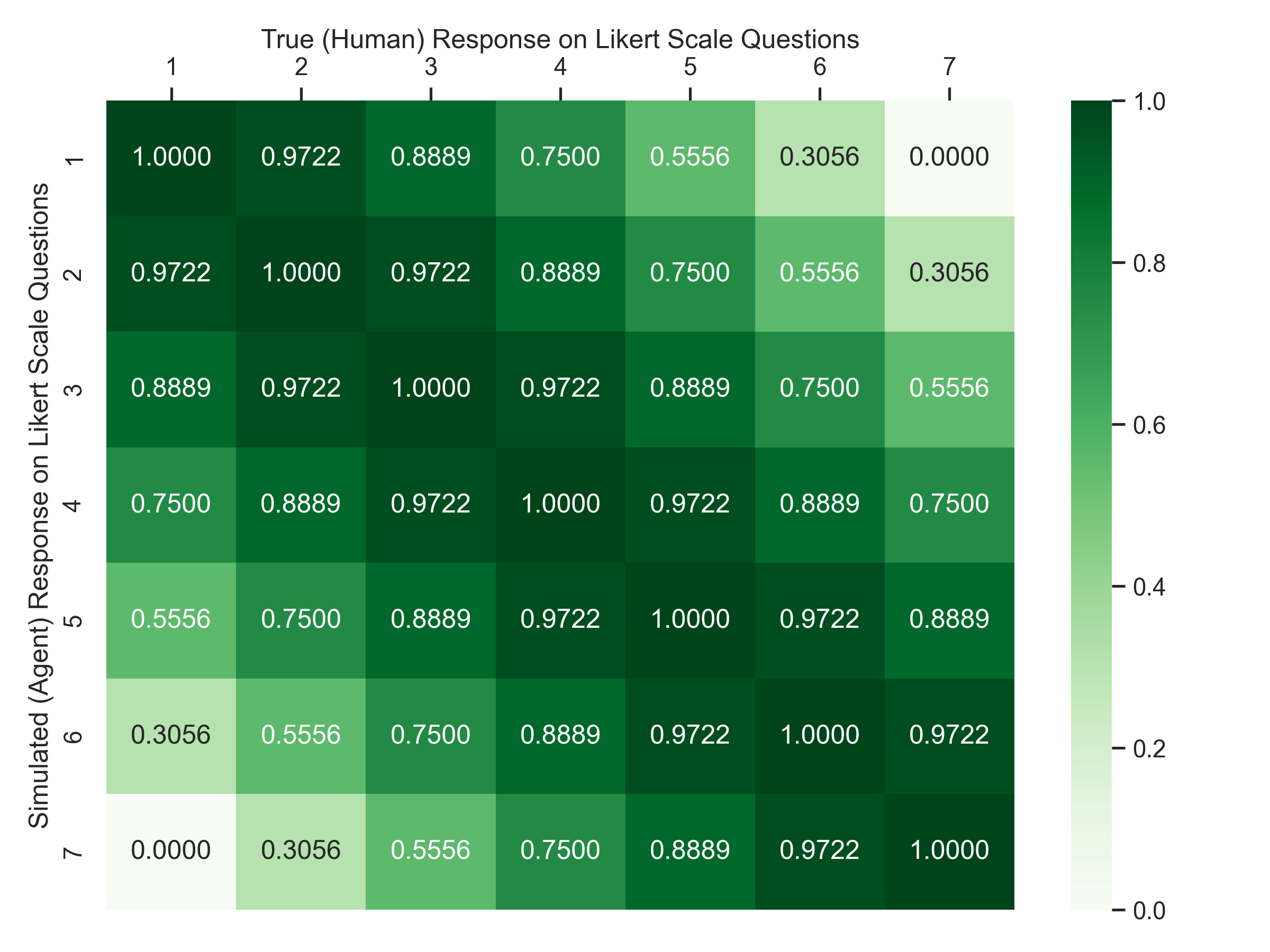}
  \caption{QWA Matrix of Simulated and Human Responses in the Agent Embodiment Task.}
  \label{fig:qwa_embodiment}
\end{figure}

\begin{figure}[htbp]
  \centering
  \includegraphics[width=0.63\textwidth]{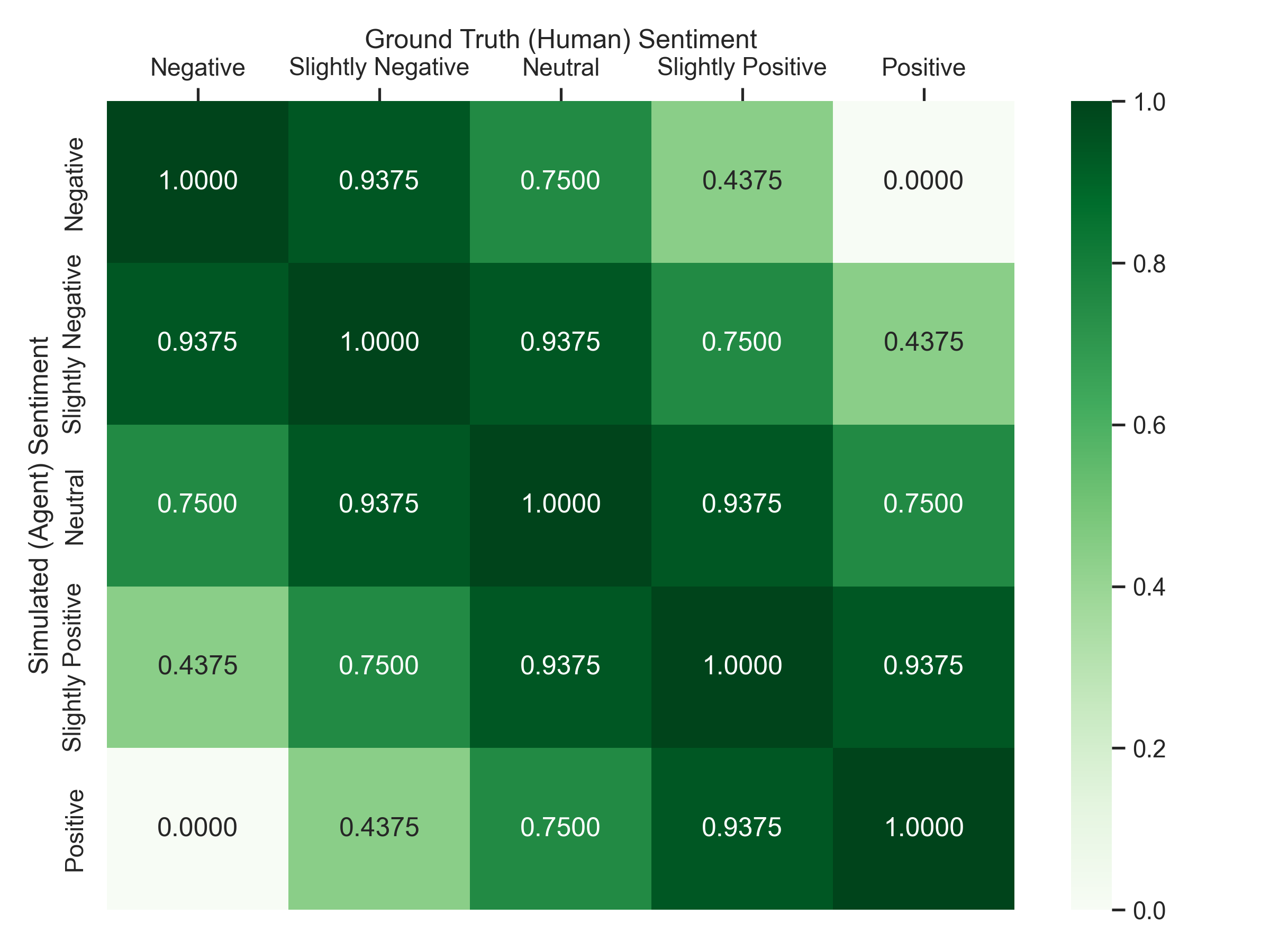}
  \caption{QWA Matrix of Simulated and Human Sentiment Responses in the Sentiment Simulation Task.}
  \label{fig:qwa_sentiment}
\end{figure}

Figures \ref{fig:qwa_embodiment} and \ref{fig:qwa_sentiment} present heatmaps of pairwise QWA scores, capturing the degree of alignment between agent-generated responses and human responses across two core tasks: agent embodiment and sentiment simulation. In both figures, each matrix cell represents the average agreement score for a specific pair of simulated and survey response values.

\supplsubsection{On Agent Embodiment Survey Replication Task}
\label{section:supp_embodiment_qwa}
Figure \ref{fig:qwa_embodiment} presents the QWA matrix for the survey replication task, where the model was prompted to generate Likert-scale responses to psychographic survey items from the perspective of an embodied agent profile. The matrix shows pairwise QWA scores between each simulated agent response (rows) and the corresponding human response (columns) on a 7-point ordinal scale.

\supplsubsection{On Sentiment Simulation Task}
\label{section:supp_sentiment_qwa}
Similarly, Figure \ref{fig:qwa_sentiment} shows the QWA matrix for the sentiment simulation task. Here, simulated sentiment responses are compared to human sentiment ratings on a 5-point ordinal scale ranging from Negative to Positive.

\supplsection{Statistical Tests}

Paired t-test was used when the assumption of normality was satisfied. The formula is given below:

\begin{equation}
t = \frac{{d}}{s_d / \sqrt{n}}
\label{eq:paired_ttest}
\end{equation}

where:

${d}$ is the mean of the differences between paired observations;

$s_d$ is the standard deviation of the differences; and

$n$ is the number of pairs.

\vspace{1em}

For group comparisons in which the normality assumption was not met, we used the Wilcoxon signed-rank test, see Equation \ref{eq:wilcoxon_signed_rank}.

\begin{equation}
W = \min(W^+, W^-)
\label{eq:wilcoxon_signed_rank}
\end{equation}

\noindent
where:

$W^+$ is the sum of positive ranks;

$W^-$ is the sum of negative ranks; and

$m$ is its sample size

\vspace{1em}

Accordingly, the Z-score formula is given below:

\begin{equation}
Z = \frac{W - \mu_W}{\sigma_W}
\end{equation}

where:

\[
\mu_W = \frac{n(n+1)}{4}
\quad \text{and} \quad
\sigma_W = \sqrt{ \frac{n(n+1)(2n+1)}{24} }
\]

$n$ is the sample size of the groups.

\vspace{1em}

In addition to hypothesis testing, we computed Cohen’s $|d|$ to quantify the effect size and assess the practical relevance of observed differences. Cohen’s $|d|$ is calculated as:

\begin{equation}
|d| = \left| \frac{\mu_1 - \mu_2}{\sqrt{ \frac{s_1^2 + s_2^2}{2} }} \right|
\end{equation}

where:

$\mu_1$ and $\mu_2$ are the means of each group; and

$s_1^2$ and $s_2^2$ are the standard deviations of each group.

\end{document}